\documentclass{icrc29}
\usepackage{graphicx,amssymb,amsmath,times}
\begin{document}
\title[Baikal neutrino telescope NT200+ ...]  
{Baikal neutrino telescope NT200+ :
Upgrade of data acquisition and time calibration systems}

\author[Aynutdinov et al ...]  { 
    V.Aynutdinov$^a$, V.Balkanov$^a$, I. Belolaptikov$^g$, N.Budnev$^b$,
    L.Bezrukov$^a$, D.Borschev$^a$,
\newauthor 
    A.Chensky$^b$, I.Danilchenko$^a$, 
    Ya.Davidov$^a$,Zh.-A. Djilkibaev$^a$, G. Domogatsky$^a$, 
\newauthor
A.Dyachok$^b$, 
    S.Fialkovsky$^d$,O.Gaponenko$^a$, O. Gress$^b$, T. Gress$^b$, O.Grishin$^b$, 
    R.Heller$^h$, 
\newauthor 
    A.Klabukov$^a$, A.Klimov$^f$, K.Konischev$^g$, A.Koshechkin$^a$, 
    L.Kuzmichev$^c$, V.Kulepov$^d$, 
\newauthor  
    B.Lubsandorzhiev$^a$, S.Mikheyev$^a$, 
    M.Milenin$^d$, R.Mirgazov$^b$, T.Mikolajski$^h$, E.Osipova$^c$,
\newauthor 
    A.Pavlov$^b$,  G.Pan'kov$^b$, L.Pan'kov$^b$, A.Panfilov$^a$,
    \framebox{Yu.Parfenov$^b$}\,, D.Petukhov$^a$,
\newauthor 
    E.Pliskovsky$^g$, P.Pokhil$^a$, V.Polecshuk$^a$, E.Popova$^c$, V.Prosin$^c$, 
    M.Rozanov$^e$, 
\newauthor 
    V.Rubtzov$^b$, B.Shaibonov$^a$, A.Shirokov$^c$, Ch. Spiering$^h$, 
    B.Tarashansky$^b$,
 \newauthor 
    R.Vasiliev$^g$, E.Vyatchin$^a$, R.Wischnewski$^h$,
    I.Yashin$^c$, V.Zhukov$^a$ \\ 
 (a) Institute for Nuclear Reseach, Russia\\
 (b) Irkutsk State University,Russia\\
 (c) Skobeltsin Institute of Nuclear Physics, Moscow State University, Russia\\
 (d) Nizni Novgorod State Technical University, Russia\\
 (e) St.Petersburg State Marine Technical University, Russia\\
 (f) Kurchatov Institute, Russia\\
 (g) Joint Institute for Nuclear Research, Dubna, Russia\\
 (h) DESY, Zeuthen, Germany}

\presenter{Presenter: R. Wischnewski (ralf.wischnewski@desy.de), \  
 ger-wischnewski-R-abs3-og27-poster}

\maketitle

\begin{abstract}

The Baikal neutrino telescope NT200, 
operating since 1998,
has been upgraded in spring 2005 to NT200+.
This telescope with 3 additional outer strings at 100~m
radius from the center
encloses a geometric volume of 5~Mtons.
We describe the modernized data acquisition and control system,
which allows for 
higher bandwidth, 
full multiplexing of data and control
streams over a single cable to shore, redundant system components
and 
underwater data pre-processing.
To calibrate all time offsets between new distant strings 
and the central telescope
on the nsec-scale, 
a new external laser unit with a powerful N2-Dye laser 
and a light diffusor, has been 
developed.
This laser 
is also used to tune reconstruction techniques for
pointlike showers with energies from 20~TeV to 10~PeV.

\end{abstract}

\section{Introduction}

The deep underwater neutrino telescope NT200+,
the  successor of the telescope NT200, started operation 
in April, 2005 \cite{IC05_stat}.
It consists of the old telescope NT200 and three external strings at 
radial distance of 100\,m from the center,
see Fig.\ref{fignt}.
NT200 was operating since 1998,  with a number of relevant physics results 
{\cite{IC05_stat,APP1,NU04,IC05_he,IC05_mon}.

The main challenge with the upgrade towards NT200+ was 
the need for a second, parallel running 
data acquisition (DAQ) and control system.
Since a simple doubling of the system was 
compatible neither  
with the number of available cable connections to shore,
nor with future 
upgrades, 
we decided to significantly modernize the 
system
by introducing for the first time embedded PCs with reliable industrial 
ethernet
infrastructure underwater.

The report describes 
(1) this DAQ upgrade and 
(2) the new laser unit, developed to calibrate 
the  time offsets for photosensors at large distances 
to nsec precision, and to simulate pointlike bright showers up to 10\,PeV.


\section{Upgrade of Data Acquisition and Control System}

The NT200 data acquisition and slow control system 
has been described in detail in \cite{APP1,KLIM}.
It is 
entirely 
based on
custom-made interfaces,
including the modems for control and for data lines to shore.
Control information is sent via the 300\,V power lines
to all modules (detector, string and  optical module-controllers),
while data are collected via a central 
modem,  
that reads out the string buffers.
For the old NT200 design, both control and data modems are located 
at shore, and require two separate underwater cables,
to avoid interferences due to cross talk.
No redundancy on this critical connection was
available.
At the shore,
experiment control was done by DOS-PCs, interfacing to
a transputer-farm which performed online event-building and
monitoring.

For NT200+, 
all data and control
cable connections 
of NT200 and the outer strings 
go through a new central control and readout unit
30\,m below surface (upper left unit in Fig.\ref{fignt})  \cite{IC05_stat}. 
At this place,
the synchronization 
between clocks of NT200 and 
the external strings (``NT+'') takes place;
it allows also for  
a centralized handling of all communications to shore.
Figure \ref{figdaq} sketches the DAQ and control system of NT200+,
composed of the two subsytems for NT200 and NT+. 
Abbreviations have the following meaning: 
{\it BED} - detector electronics module, 
{\it BEG} - string electronics  module, 
{\it BSD} - module to measure NT200/NT+ relative trigger times.
Trigger formation is by BED for NT200 and by BEGs for NT+ \cite{IC05_stat}.
The slow control connection to all 8 and 3 strings, respectively, is done
via electronic fanout units ({\it ``Relay''}).

For the new system, 
we kept all front-end units and the internal telescope buses unchanged, 
and
introduced two central ``underwater PC spheres'' 
(see Figs.\ref{figdaq} and \ref{figbai-daq1}).  
They are housing the experiment data and control modems
for NT200 and for the external strings,
formely located at shore.
The modems are handled by two embedded Linux PCs 
in PC/104 standard,
which are 
connecting 
via tcp/ip sockets  
to the shore data center PCs
for NT200 and NT+. 
The connection to shore is by a 
single
DSL-Modem 
at a speed of up to 2Mbit/s.
This 
full multiplexing of all  
data and control streams (NT200 and NT+)
reduces the number of shore wires to two, 
and allows for a spare connection, 
which is in ``cold standby'' mode
(activated only in case of failure, 
or if larger bandwidth 
would be needed).

Both ``underwater PC spheres'' are nearly identical,
their content 
is detailed
in Fig.\ref{figbai-daq1}: 
a single board PC/104 ({\it PC104:}  Advantech-PCM9340), 
a DSL-modem 
({\it DSL-M:} FlexDSL-PAM-SAN, with hub and router),
a managed 
ethernet switch 
({\it SwRSTP:} RS2-4R, running RSTP 
protocols for the two-fold redundant  
ethernet network between the PC spheres), 
an Ethernet-ComServer 
({\it CSrv:} WUT-58211, for PC-terminal emulation), 
two media-converters  
({\it Mc:} for coaxial connection to external control units)
and the experiment data and control modems ({\it D-Mod} and {\it C-Mod}).
Both PC spheres are connected via two twisted pair cables 
(100Base-TX).
This underwater system works stable since its 
first installation in 
2004.
The old shore transputer farm was replaced by a 
powerful
Linux PC.
Using Linux throughout the system allows for 
easy remote control
from home institutions.

Modern off-the-shelf data communication and automatization 
components  were used 
for all new components and functionalities.
This way a modern, reliable system with much improved performance 
could be built and commissioned
with minimal effort.
The system is scaleable to 
modifications of the detector design,
and 
is ready for
physics performance improvement by
e.g. introduction of more complex trigger strategies
(low-level trigger decisions with underwater PCs).
We are also considering 
the possibility to 
upgrade 
components like string and detector controllers, 
to improve 
physics potential 
of  NT200+.


\begin{figure}[htbp]
\begin{minipage}[b]{.48\linewidth}
\centering\includegraphics*[width=0.7\textwidth,angle=0,clip]{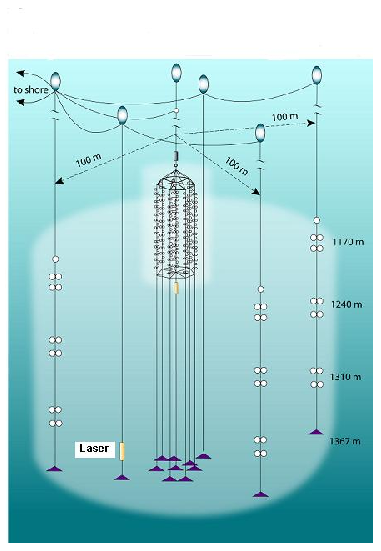}
\caption{
NT200+: the old NT200 telescope surrounded 
by 3 new strings at 100\,m radius.
The new NT200+ laser is indicated; 
central DAQ is located in left upper unit.  
}
\label{fignt}
\end{minipage}
\hfill
\begin{minipage}[b]{.48\linewidth}
\centering\includegraphics*[width=0.7\textwidth,angle=0,clip]{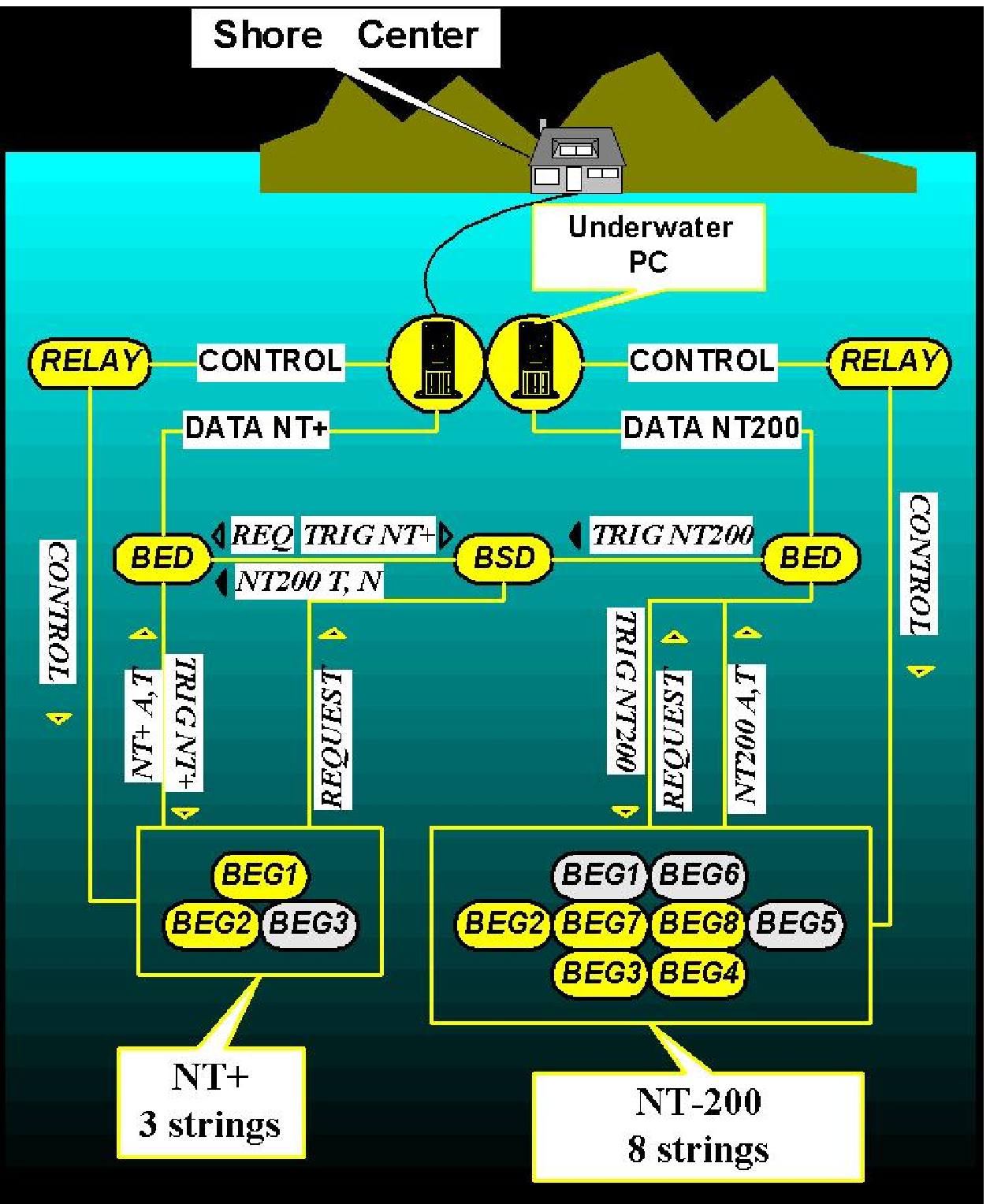}
\caption{
Sketch of  data collection and slow control in NT200+: 
8-string telescope NT200 and 
3 new outer strings (``NT+''),
controlled from two underwater PCs
(see Fig.\ref{figbai-daq1}).  
}
\label{figdaq}
\end{minipage}
\end{figure}

\begin{figure}[htbp]
\begin{minipage}[b]{.50\linewidth}
\vspace*{-3mm}
\includegraphics*[width=1.3\textwidth,angle=0,clip]{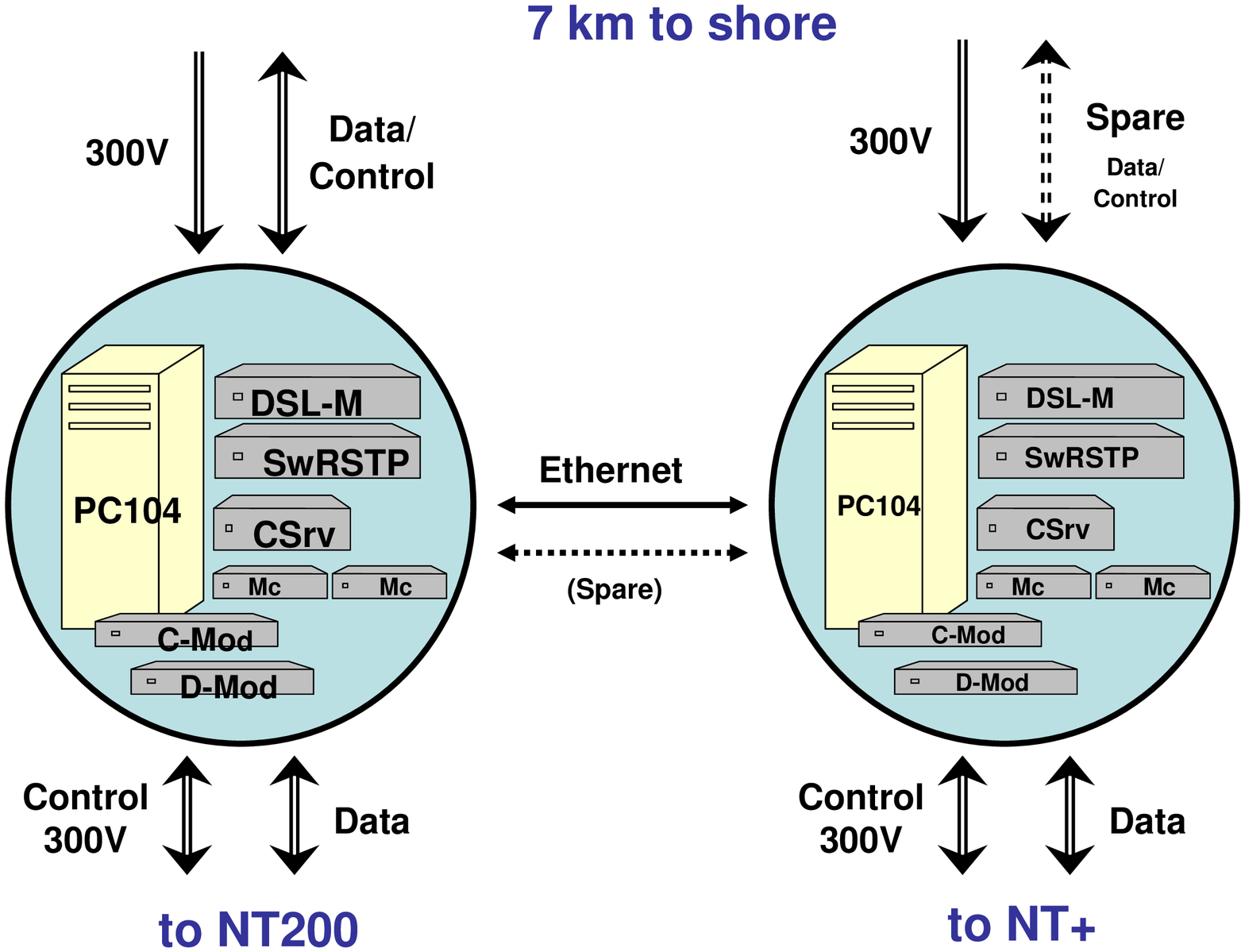}
\caption{
Sketch of the new NT200+ central DAQ/Control spheres
with embedded PC104, 
redundant 
network components, 
and the experiment bus interfaces.
See also Fig.\ref{figdaq}.
}
\label{figbai-daq1}
\end{minipage}
\hfill
\begin{minipage}[b]{.45\linewidth}
\centering\includegraphics*[width=0.9\textwidth,angle=0,clip]{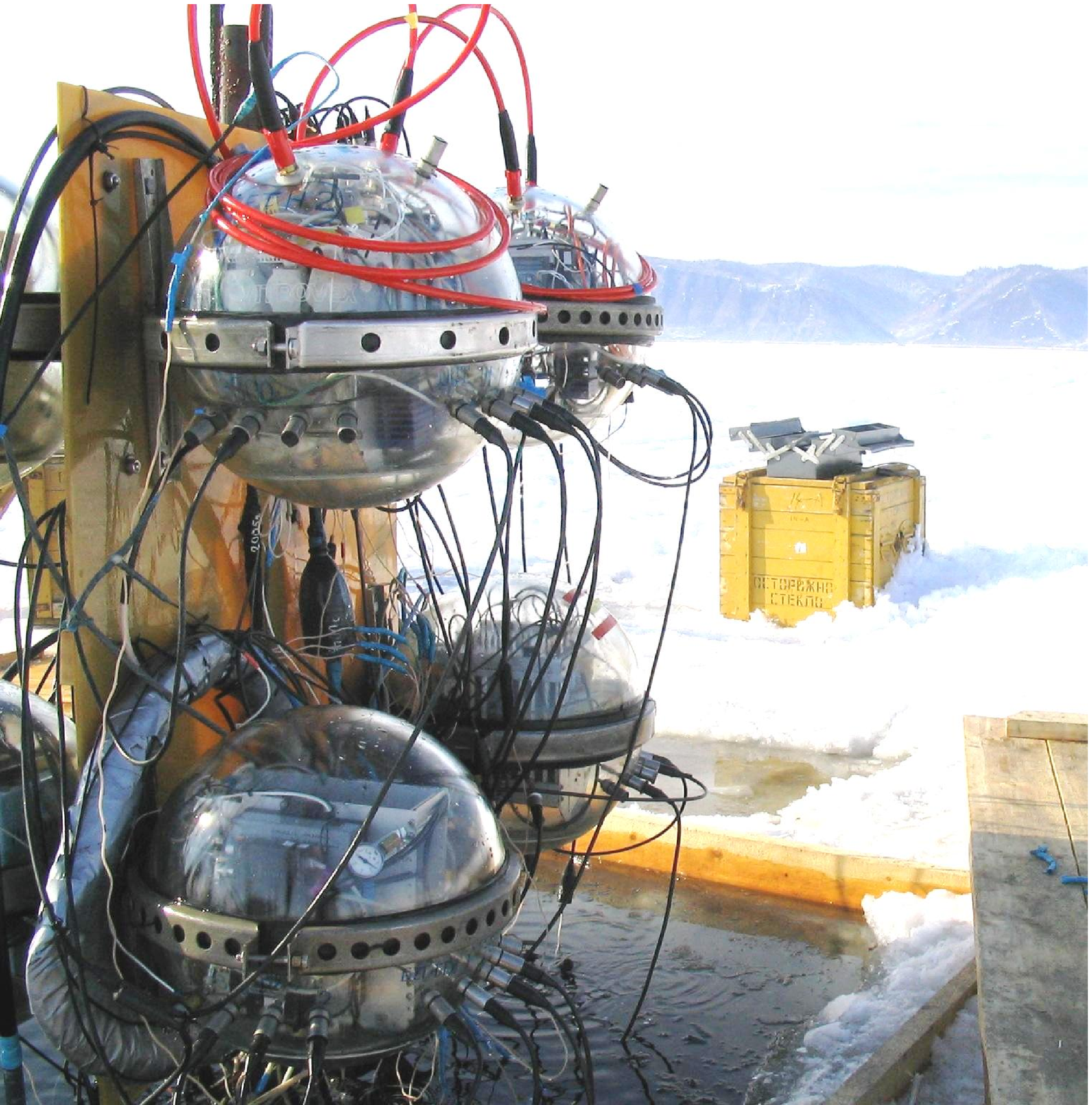}
\caption{
NT200+ central DAQ electronics 
in its pressure housings, before deployment. 
Upper spheres are underwater PC-NT200
and PC-NT+, 
detailed in Fig.\ref{figbai-daq1}.}
\label{foto_3}
\end{minipage}
\end{figure}


\section{NT200+ Laser}

Large volume underwater Cherenkov detectors need
calibration of
the relative time-offsets between all light-sensors
to a precision of a few nsecs, since 
event reconstruction and classification
are based 
on the precise light arrival times.
For NT200+, interstring calibration 
(and for outer strings also intrastring calibration) 
is done with a
short and powerful external laser light source
with $>10^{12}$ photons per pulse and nsec-pulse duration,
located between two outer strings and close to lake bottom,
see Fig.\ref{fignt}.
This ensures amplitudes of 
$\sim$100
photoelectrons on a few 
PMTs on each external string and on NT200.
High amplitudes
minimize 
uncertainties
due to light scattering.
If the emission characteristics is also 
isotropic, 
the light source can be used to imitate light and amplitude
patterns from high energy particle cascades 
and to verify energy and vertex reconstruction
\cite{NU04}.

The NT200+ laser calibration unit is made of a 
powerful
short-pulse Nitrogen laser
($\lambda = 337$\,nm) with about 100\,$\mu$J for $<$1\,nsec pulse duration.
It is pumping a Coumarin dye laser at 480\,nm, 
which yields about 10\% of the original intensity.
After passing through a computer-controlled attenuator disc
the light is isotropized 
by a light diffuser ball, made of a 
round-bottom flask filled with Silicone Gel (RTV-6156) 
admixed with hollow micro-glass spheres at about 5\%-volume ratio
(S32 from 3M, with $\approx 40\mu$m diameter;
following an idea developed for SNO \cite{SNO}).
The total loss of this isotropizing sphere 
is $\approx$ 25\%. 
This guarantees that light output at maximum intensity is
well above the design value of $> 10^{12}$ photons/pulse.

All components are mounted into a 1\,m-long cylindrical glass 
pressure housing, which gives  
isotropic emission for more than the upper hemisphere.
The unit is 
installed at a depth of 1290\,m below surface and
operated in autonomous mode: after power-on from shore,
a series 
of pulses
at various intensities is conducted. 

This laser allowed an 
independent performance check of 
the
key elements of the NT200+ timing system. 
We performed the relative time synchronization
of all news strings and NT200, and  find the jitter of this 
to be less than 3\,nsec (see also \cite{IC05_stat}).
The laser unit will be used, varying the total intensity, 
to calibrate pointlike shower vertex and 
energy reconstruction algorithms for 
energies from 20\,TeV to 10\,PeV.


\section{Conclusions}

The Baikal neutrino telescope NT200 was upgraded 
to the much larger detector NT200+ in spring 2005, 
by adding three distant outer strings.
The modernized data acquisition and control system 
uses 
for the first time 
PCs underwater, 
interconnected 
by a redundant 
ethernet network.
This allowed 
to unify the
connections from shore station to the underwater site:
all data and control lines are multiplexed 
via a single DSL-line from shore. 
For all new components and functionalities
modern off-the-shelf data communication and automatization 
components  were used,
facilitating to build a
reliable and scalable system 
with minimal 
effort.
The new system can be viewed as a test-bed for 
a larger (km$^3$) detector project.

A new external laser unit was constructed 
for calibration of 
time 
offsets between 
optical modules 
over the large distances in NT200+.
This 
nsec-pulse N2/Dye laser with a light-isotropizer 
and 
intensity of $>\,10^{12}$ photons/ pulse 
is 
also used to 
simulate 
point-like showers up to 10\,PeV.
We verified that for 
the NT200+ 
detector design, 
based 
on electrical (coaxial) cables of up to km's,
an overall time synchronizaton 
of $<$3\,nsec 
can be reached.

\vspace*{-1.5mm}

\section{Acknowledgements}

{\it This work was supported by the Russian Ministry of Education and Science,
the German Ministry of Education and Research and the Russian Fund of Basic 
Research} ({\it grants} \mbox{\sf 05-02-17476}, \mbox{\sf 04-02-17289} 
{\it and} \mbox{\sf 02-07-90293}), {\it by the Grant of President of 
Russia} \mbox{\sf NSh-1828.2003.2}, {\it and by NATO-Grant} 
\mbox{\sf NIG-981707}.


\end{document}